# Layer-selective non-reciprocal electric-field switching of magnetism in van der Waals heterostructure multiferroics

Yangliu Wu[1,2], Deju Zhang[3], Yanning Zhang[3]✉, Longjiang Deng[1,2]✉ and Bo Peng[1,2]✉

[1]National Engineering Research Center of Electromagnetic Radiation Control Materials, School of Electronic Science and Engineering, University of Electronic Science and Technology of China, Chengdu, 611731, China
[2]Key Laboratory of Multi Spectral Absorbing Materials and Structures of Ministry of Education, University of Electronic Science and Technology of China,
Chengdu 611731, China
[3]Institute of Fundamental and Frontier Sciences, University of Electronic Science and Technology of China, Chengdu 611731, China
✉Correspondence author's e-mail: bo_peng@uestc.edu.cn; denglj@uestc.edu.cn; yanningz@uestc.edu.cn

**Abstract: The search of two-dimensional (2D) multiferroic materials, particularly the coexistence of ferromagnetic and ferroelectric orders, is exciting yet remains challenging. Here, we report a van der Waals (vdW) heterostructure multiferroic comprising atomically thin layered antiferromagnet (AFM) $CrI_3$ and ferroelectric (FE) $\alpha\text{-}In_2Se_3$. We demonstrate anomalously layer-selective nonreciprocal and nonvolatile electric-field control of magnetization by the ferroelectric polarization. The nonreciprocal electric control originates from an intriguing antisymmetric enhancement of interlayer ferromagnetic coupling in the opposite ferroelectric polarization configurations of $\alpha\text{-}In_2Se_3$, which favor to selectively switch the spins in the second layer. Our work provides numerous possibilities for creating diverse heterostructure multiferroics at the limit of few atomic layers for multi-stage magnetic memories and brain inspired in-memory computing.**

Switching magnetism using pure electric fields instead of magnetic fields[1] in storage mediums is a promising approach for next-generation low-power spintronic and memory devices[2-5]. Magnetoelectric coupling, particularly in multiferroic systems, opens the door for the manipulation of digital information by utilizing the cross-coupling between magnetic and ferroelectric orders[6-9]. Unfortunately, single-phase multiferroic materials are rare due to the inherent mutual exclusion between magnetism and electric polarization, and the magnetoelectric coupling is usually weak[10]. Several works have been reported to control magnetization in multiferroic heterostructures film through indirect strain-mediated coupling[11,12]. However, reversible non-volatile magnetization rotation via direct magneto-electric coupling is a major challenge and encounters unprecedented dilemmas.

In van der Waals heterostructure systems, the interlayer coupling between adjacent atomically thin layers renders the electronic wave function to reside in more than one specific layer, promoting the layer degree of freedom from a spatial variable index to a quantum mechanical index. The recent discoveries of 2D vdW magnets[13-17], and ferroelectrics[18-23], have attracted much attentions. These atomically thin 2D crystals provide new avenues for studying the interface coupling between magnetic and ferroelectric orders through vdW engineering[24,25]. Switching ferroelectric polarization consumes three orders of magnitude less energy than switching ferromagnets[8]. Moreover, ferroelectric polarization can modulate the overlapping electronic wave function of distinct layers by the electric field, leading to significant interfacial magneto-electric coupling effect and thereby efficiently manipulate the magnetic properties in a non-volatile fashion. An exciting route towards low-power and non-volatile spintronics would be to integrate the ferroelectric polarization and magnetic orders for creating a 2D FM/FE heterostructure, and for directly switching spin by the pure electric field, beyond the limit of weak magneto-coupling in the single-phase multiferroics. 2D vdW FM/FE multiferroic heterostructures enable functions beyond the reach of isolated materials, and open new avenues for realizing non-volatile electric control of magnetism. Although, several theoretical works have proposed to realize the magnetic phase transition by changing the direction of ferroelectric polarization in a 2D vdW FM/FE heterostructure multiferroic[26,27]. However, the experimental manipulation of magnetism through ferroelectric polarization has rarely been reported and remains challenging in a 2D vdW heterostructure multiferroic, in particular nonvolatile magnetization reversal upon a small voltage.

Here we report a multiferroic heterostructure that consists of 2D layered AFM $CrI_3$ and FE α-$In_2Se_3$. Reversing the FE polarization direction of the α-$In_2Se_3$ nanoflake switches the interlayer magnetism. The opposite FE polarization ***P*** configurations create a different interlayer coupling, which results in a nonreciprocal magnetoelectric coupling effect in the transitions between two magnetic states. Furthermore, when the electric field is removed, the residual FE polarization can shelter the magnetoelectric coupling, inducing a non-volatile electric control of magnetism. This work demonstrates unique and rich physical phenomena originating from 2D AFM/FE interfaces, and opens a way to the continuously electric manipulation of magnetism toward ultralow-power spintronics with non-volatility.

## VdW multiferroic heterostructure and anomalous interface coupling

The vertically stacked multiferroic heterostructures comprise a 2D layered antiferromagnet $CrI_3$ and a FE $\alpha$-$In_2Se_3$ nanoflake, sandwiched between two graphene electrodes (Fig. 1a). To study the mechanism and limit of interfacial electromagnetic coupling effect and weaken the influence of stress, a six-layer AFM $CrI_3$ was chosen. A middle hBN flake serves as a good insulating medium to block leakage current and guarantee the remanent FE polarization in $\alpha$-$In_2Se_3$. Figure 1b presents a micrograph of a sandwich multiferroic device. Even layer $CrI_3$ is a layered antiferromagnet[15]. As shown in Fig. 1c, the magnetic moments of $Cr^{3+}$ cations are aligned in each layer, while their interlayer coupling is antiferromagnetic; this leads to a vanishing magnetization at zero magnetic field. The sketch of side-view and top-view for crystal structures and atomistic arrangements of $\alpha$-$In_2Se_3$ are shown in Fig. 1d. The central selenide (Se) layer is tetrahedrally encapsulated by two neighboring indium (In) atoms, whereas the Se atoms at the top and bottom surface layer regularly fill in the side hollow sites of the In atoms. As shown in Fig. 1d (top panel), the "ABC" atomic arrangement gives a hexagonal in-plane projection, in which Se and In atoms are arranged in a triangular lattice at the six vertexes, respectively. The $\alpha$-$In_2Se_3$ is constructed by dislocated quintuple layer, leading to spatial inversion symmetry breaking. It is demonstrated that an out-of-plane FE polarization in hexagonal $\alpha$-$In_2Se_3$ always exists regardless of flake thickness[28,29]. As indicated in Fig. 1e, the Raman modes are observed near 94, 159, 179, and 256 $cm^{-1}$, which can be attributed to the $E^2$, $E^3$, $E^4$, and $A_1^4$ modes of $\alpha$-$In_2Se_3$[30,31], respectively. Four Raman features appear at 77, 108, 129, and 240 $cm^{-1}$, labeled inverted triangle in Fig. 1e, which are assigned to the $A_g$ and $E_g$ modes, of 2D magnet $CrI_3$, respectively[32,33]. All Raman and magneto-optical measurements were carried out at 10 K in a closed cycle He cryostat.

The out-of-plane magnetization was probed by reflective magnetic circular dichroism (RMCD) microscopy on the multiferroic heterostructure devices (see Methods for details). The laser energy at 1.96 eV (633 nm) was chosen which nears the absorption edge of $CrI_3$[15]. The RMCD signal is proportional to the sample's out-of-plane magnetization. Figure 1f shows the RMCD signal of the six-layer $CrI_3$ on hBN and FE $\alpha$-$In_2Se_3$ as a function of the out-of-plane magnetic field, respectively. The yellow curves correspond to sweeping magnetic fields from a positive to a negative value, and the green curves shows the inverse process. As the magnetic field sweeps, the RMCD signals show seven plateaus, which indicates six layers of $CrI_3$ flake. For $CrI_3$ on hBN (Position indicated by red dot in Fig. 1b), the RMCD signal nearly vanishes between -0.9 to +0.9 T, corresponding to the interlayer AFM states, with degenerated state of ↑↓↑↓↑↓ or ↓↑↓↑↓↑ (1-6 layers from bottom to top. Represented by S0 in Fig. 1f), where the arrows indicate the out-of-plane magnetizations. A remnant

RMCD signal originates from the uncompensated AFM interlayer coupling[15]. As the positive magnetic field increases, the RMCD signals show three plateaus. Using first-principles calculation (Supplementary Fig. 1), we identify the corresponding magnetic states of these three plateaus as S1$_+$ (↑↓↑↓↑↑ or ↑↑↓↑↓↑), S2$_+$ (↑↑↑↓↑↑), and S3$_+$ (↑↑↑↑↑↑), respectively. For the sake of description, S1 temporarily represents only one magnetic state (↑↓↑↓↑↑) throughout the article. There are also three magnetic states for the negative field plateaus: S1$_-$, S2$_-$ and S3$_-$, which are the three time-reversal copies (S$_+$) of the positive field counterparts. This spin-flip behavior of six-layer CrI$_3$ on hBN is consistent with the reported result[34].

We next study the CrI$_3$ on FE α-In$_2$Se$_3$ (position indicated by yellow dot). The RMCD vs out-of-plane magnetic field are measured upon sequentially applied -10V, 0V and +10 V, respectively (top three curves in Fig. 1f). The spin-flip magnetic fields ($\mu_0H_{sf}$) of S0$_\pm$→S1$_\pm$ and S2$_\pm$→S3$_\pm$ magnetic states with FE polarization from α-In$_2$Se$_3$ are smaller than that on hBN (Fig. 1f and Supplementary Fig. 2a, b, c), which suggest that the FE polarization enhances the interlayer ferromagnetic coupling. Importantly, a non-reciprocal and selective electric control of magnetism takes place on S1$_\pm$→S2$_\pm$, in which only the spins of the second layer are selectively manipulated. The $\mu_0H_{sf}$ of S1$_+$→S2$_+$ upon downward FE polarization (***P*** down, +10 V) is 0.101 T smaller than that without FE polarization in the case of CrI$_3$ on hBN, whereas the $\mu_0H_{sf}$ of S1$_+$→S2$_+$ under upward FE polarization (***P*** up, -10 V) remain approximately unchanged (Supplementary Fig. 2 and 4).

**Non-reciprocal magnetoelectric coupling**

The first-principles calculations suggest that flipping the ferroelectric polarization of α-In$_2$Se$_3$ from upward to downward modulates the charge distribution at the I-I interface space of CrI$_3$/In$_2$Se$_3$, modulating interlayer exchange coupling of CrI$_3$. By comparing the free energy of different magnetic ground states in the two polarization configurations, it is found that the energy difference of adjacent magnetic states decreases in both downward and upward polarization configurations (Supplementary Fig. 2d). The decreasing values of energy difference for the S1$_+$→S2$_+$ in ***P*** down configuration (+10 V) is about 30 times larger than that in ***P*** up configuration (-10 V), whereas the others are reduced by the same order of magnitude in both downward and upward polarization configurations (Supplementary Fig. 2d). This behavior is good agreement with the FE polarization dependence of the $\Delta\mu_0H_{sf}$ for the state transition (red marker in Supplementary Fig. 2d). This suggests that the selectively non-reciprocal electric control of magnetism originates from an intriguing antisymmetric enhancement of interlayer ferromagnetic coupling in the opposite ferroelectric polarization configurations of α-In$_2$Se$_3$. Furthermore, the Raman features of 2D CrI$_3$ remain unchanged under different voltages (Supplementary Fig. 3), suggesting the absence of strain-mediated magnetoelectric coupling[35], nevertheless which are common in conventional multiferroic heterostructures[11,12].

We further study the electrical control behaviors of the $CrI_3$/$In_2Se_3$ heterostructure multiferroic device by sweeping the gate voltages ($V_g$) from -10 V to +10 V. Figure 2 presents the maps of normalized RMCD intensity as a function of the applied voltage and magnetic field (sweeping up, $-H \rightarrow +H$; sweeping down, $+H \rightarrow -H$). In the $V_g$-$\mu_0H$ dependent magnetic phase diagram, the light color represents low RMCD signal with a weak magnetization arising from uncompensated layered AFM coupling (mark with $S1_+$, $S2_+$, $S1_-$, and $S2_-$), and the dark red and blue colors represent the two strong FM states (fully spin-polarized states) with opposite signs (mark with $S3_+$ and $S3_-$). The maps of normalized RMCD intensity sweeping up from -0.8 T to +0.8 T as a function of the external voltage show weaker magnetizations (Supplementary Fig. 5.). Based on this magnetic phase diagram, the magnetic phase boundaries, that is, the spin-flip field ($\mu_0H_{sf}$) for the magnetic transitions, strongly depends on the gate voltage. Strikingly, this gate voltage dependent behavior only occurs during a transition from a low energy magnetic state to a high magnetic energy state with increasing magnetic field (Fig. 2 and Supplementary Fig. 6). The energy barrier induced by ferroelectric polarization is not enough to form a stable intermediate magnetic state, which may be the reason it is difficult to observe electrically controlled magnetism during the transition from high energy magnetic state to low energy magnetic state. In Fig. 2e, we look closer at the magnetic transition by comparing the voltage dependence of spin-flip magnetic field. The $\mu_0H_{sf}$ exhibits a maximum value near $V_g = +2$ V instead of 0 V, indicating a residual ferroelectric polarization at 0 V and a coercive voltage of ~2 V. The shift in $\mu_0H_{sf}$ for magnetic transition $S1_- \rightarrow S2_-$ is unidirectional and non-linear. Remarkably, the spin-flip field $\mu_0H_{sf}$ for magnetic transition $S1_- \rightarrow S2_-$ has roughly shifted from -1.52 T to -1.65 T by a significant change of 0.13 T at +10 V, in stark contrast, the $\mu_0H_{sf}$ is maintained nearly with no change at -10 V (Fig. 2e); it is strongly non-reciprocal electrical control of magnetism (for the time-reversal counterpart at negative magnetic fields, see Fig. 2f). This is consistent with strong nonsymmetric magnetoelectric coupling at two opposite polarization configurations based on first-principles calculations (Supplementary Fig. 2d).

## Layer-selective electrical manipulating of spin rotation

Due to this strong dependence of the spin-flip field on the gate voltage, we can realize an electrical control of transition between the $S1_-$ ($S1_+$) and $S2_-$ ($S2_+$) phases. Figure 3a and b show such an electrically controlled RMCD curves at various FE polarization configurations, and corresponding to counterpart at negative magnetic fields (horizontal line cuts of the phase map of Fig. 2b and d). For the convenience of comparison, the RMCD intensity is normalized from 0 (S1_ or S1_+) to ±1 (S2_ or S2_+). The RMCD signals are near zero in the range from +1.5 (-1.5) to +1.6 T (-1.6 T) at 0 V, corresponding to magnetic state S1_+ or S1_-. As the voltage increases to +10 V, the RMCD signal gradually rises and reaches a saturated value (±1), demonstrating a 180° spin reversal (mark with red dashed boxes, Supplementary Fig. 6b and d). However, as the voltage increases to -10 V, the RMCD signals only show a slight enhancement, indicating spin canting rather than reversal. Fig. 3c and d show RMCD signals as a function of voltage at four selected magnetic fields, which shows a continuous manipulation of spin rotation in the vdW multiferroics via electrical gating. Fig. 3e and f show $\mu_0H$-$V_g$ dependent absolute value difference of RMCD, where RMCD curve at 0 V serves as a reference. For the magnetic fields ranging from +1.50 T (-1.50 T) to +1.63 T (-1.63 T), the required voltage for initiating the magnetic transition from the S1_+ (S1_-) to the S2_+ (S2_-) shifts towards positive gate voltage, along with decreasing applied magnetic field.

## Non-volatile performance

More importantly, to checking the non-volatile electrical control of magnetic transition, we performed the time-dependent RMCD measurements after remove applied gate voltage. First, a -10 V voltage is applied to recover to the initial magnetic state as shown in the dark blue line in Fig. 4a and b, and then a +10 V voltage is applied to initiate the transition from the S1_+ (S1_-) to the S2_+ (S2_-). Fig. 4a and b show time-dependent RMCD curves from 0 to 24 h after removing the +10 V gate voltage, which show that the electric-tuned RMCD curves still have been persist and do not return to the initial magnetic state even after 24 hours. To quantitatively understand the non-volatile electrical control of magnetic transition between S1_+ (S1_-) and S2_+ (S2_-), we analyzed the magnetic-field dependent RMCD difference as a function of time, where the RMCD curve of the initial magnetic state serves as a reference. Evidently, above 50% of the RMCD intensities has been preserved ranging from +1.64 T (-1.640 T) to +1.68T (-1.68 T) after 24 hours, indicating a non-volatile electric control of magnetic transition (mark with black dashed boxes, Fig. 4c and d). Thus, the $CrI_3$/$In_2Se_3$ heterostructure multiferroic is a promising magnetoelectric functional material with the potential for multistate storage, realizing a controllable magnetic reading and electric-field writing function.

## Conclusions

We have demonstrated layer-selective non-reciprocal and non-volatile electric control of magnetism in a van der Waals $CrI_3/In_2Se_3$ heterostructure multiferroic. The critical field for the magnetic transition can be tuned by up to 0.13 T by applying a small gate voltage, allowing a gate-voltage-driven continuous magnetic phase transition ($S1_+$ ($S1_-$) to $S2_+$ ($S2_-$)). This electric-field control of magnetic transition is reversible due to the robust non-reciprocal magnetoelectric coupling effect induced by opposite polarization configuration. The ferroelectric polarization modulation of interlayer exchange coupling in 2D $CrI_3$ leads to non-volatile electrical switching between two magnetization states, which could open a way to a new class of ultralow-power and multistate spintronic devices.

## Methods

### Sample fabrication

$CrI_3$, α-$In_2Se_3$, hBN and graphene nanoflakes were mechanically exfoliated from bulk crystals via PDMS films in a glovebox, which were bought from Chengdu Mouton tech companies. All exfoliated α-$In_2Se_3$, hBN, $CrI_3$ and graphene flakes were transferred onto pre-patterned Au electrodes with thickness of 50 nm on $SiO_2$/Si substrates one by one to create multiferroic heterostructure in glovebox and were further in-situ loaded into the cold head for optical measurements.

### Optical measurements

The Raman signals were recorded by a Witec Alpha 300R Plus low-wavenumber confocal Raman microscope. For Raman measurements, a 633 nm laser of ~2 mW is parallel to the X-axis (0°) and focused onto samples by a long working distance 50× objective (Zeiss, NA = 0.55).

The RMCD measurements were performed based on the Witec Alpha 300R Plus low-wavenumber confocal Raman microscope, coupled with a magneto-optical system (Mottec MO-100, Chengdu Mouton tech companies), a closed cycle superconducting magnet (7 T) and a closed cycle He optical cryostat (10 K) with an electronic transport measurement system. A free-space 633 nm laser of ~3 μW modulated by photoelastic modulator (50 KHz) was reflected by a non-polarizing beamsplitter cube (R/T = 30/70) and then directly focused onto samples by a long working distance 50× objective (Zeiss, NA = 0.55). The reflected beam which was collected by the same objective passed through the same non-polarizing beamsplitter cube and was detected by a photomultiplier tube, which was coupled with lock-in amplifier and Witec photocurrent scanning system. The magnetic field and gate voltage were applied by the superconducting magnet and 155 voltage sources.

## Data availability

The data that support the plots within this paper are available from the corresponding authors upon reasonable request. Source data are provided with this paper.

## Code availability

The codes used for plotting the data are available from the corresponding author upon reasonable request.

**Acknowledgements** B.P., and L.D. acknowledge support from National Science Foundation of China (52021001). B.P. acknowledge support from National Science Foundation of China (62250073).

**Author contributions** B.P conceived the project. Y.W. prepared the samples and performed the magneto-optical and Raman measurements assisted by B.P., and analyzed and interpreted the results assisted by Y.Z., L.D. and B.P.. J.D.Z. and Y.Z. performed the first-principles calculations. Y.W. and B.P. wrote the paper with input from all authors. All authors discussed the results.

**Competing interests**
The authors declare no competing interests.

**Additional information**
**Supplementary information** The online version contains supplementary material available at https://

**Correspondence and requests for materials** should be addressed to Bo Peng.

**Peer review information** *Nature Electronics* thanks xxxx and the other, anonymous, reviewer(s) for their contribution to the peer review of this work.

**Reprints and permissions information** is available at http://www.nature.com/reprints.

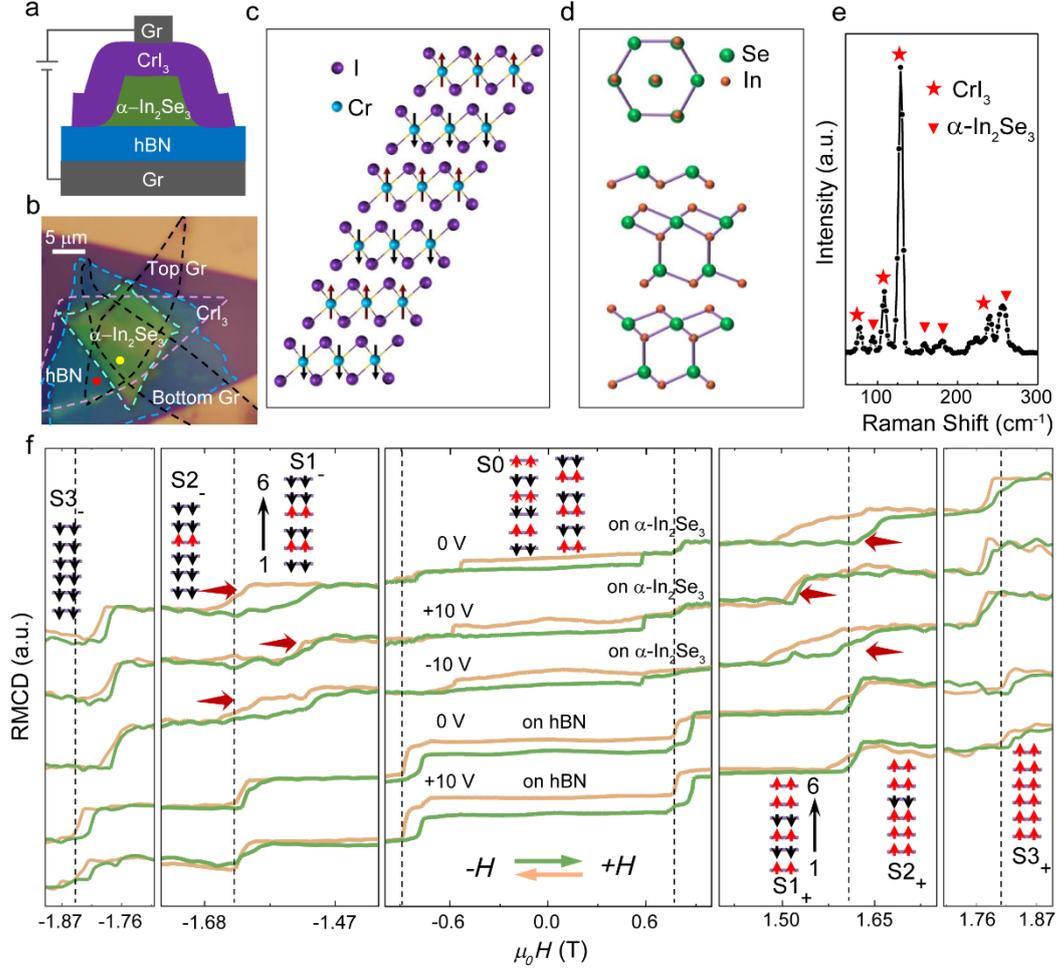

**Fig. 1 | CrI$_3$/ In$_2$Se$_3$ heterostructure multiferroic device and magnetic orders of six-layer CrI$_3$ at 10 K**. **a,** A schematic side view of multiferroic device sandwiched between graphene contacts and hBN. **b,** Optical micrograph of device. **c,** View of the in-plane atomic lattice of six-layer CrI$_3$. Cr$^{3+}$ and I$^-$ ions are shown as blue and purple balls, respectively. Six-layer CrI$_3$ shows a layered antiferromagnet when cooled below the Néel temperature. **d,** Atom structure models for the top and side views of hexagonal α-In$_2$Se$_3$, indicating a non-centrosymmetric structure. **e,** Raman spectra of multiferroic heterojunction at 10 K. **f,** RMCD curves of six-layer CrI$_3$ on α-In$_2$Se$_3$ and hBN at selected electric field. The arrows represent the spin direction and show the magnetic states of 1-6 layers from bottom to top (S1$_+$: ↑↓↑↓↑↑).

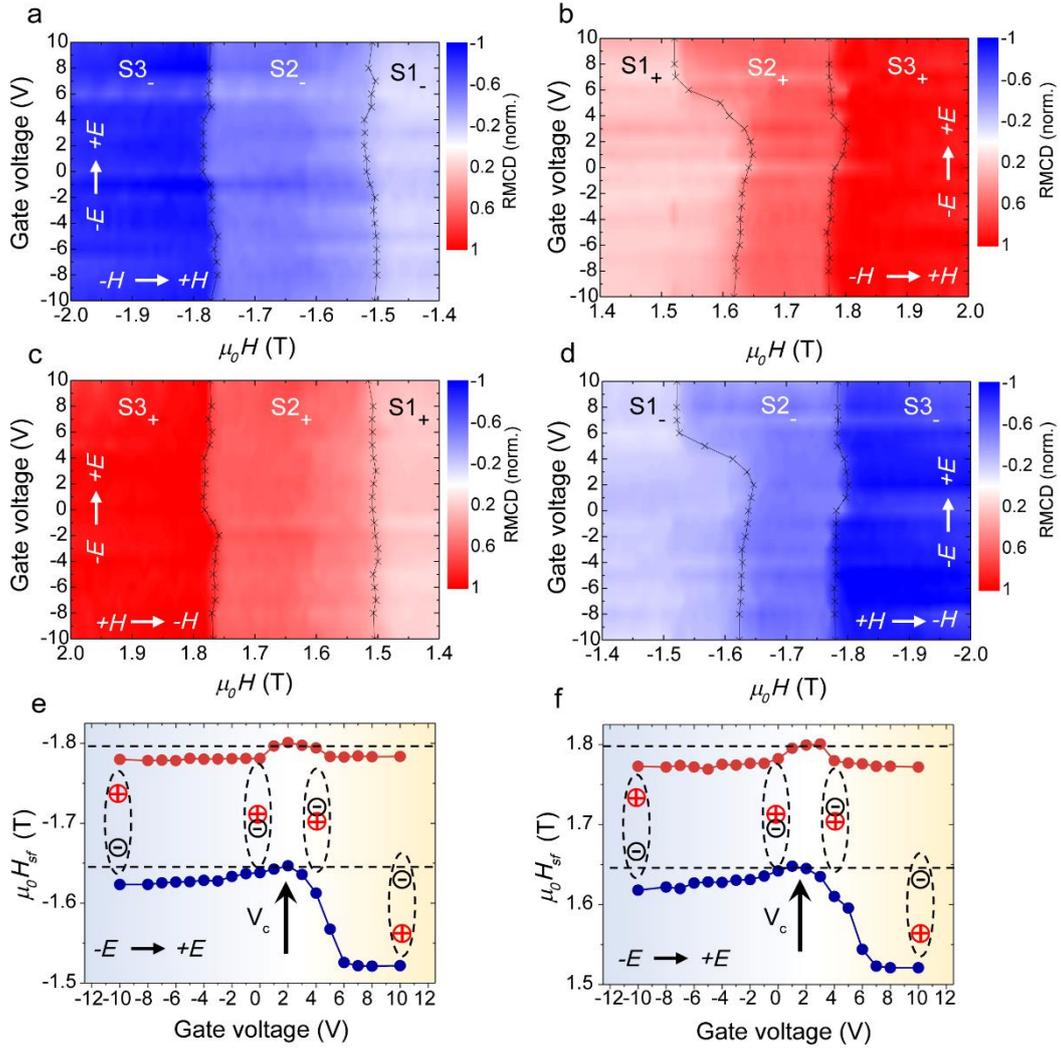

**Fig. 2 | Antisymmetric magnetoelectric coupling effect in GrI$_3$/In$_2$Se$_3$ multiferroic heterojunction. a, b,** RMCD intensities of a multiferroic device as a function of both applied gate voltage and magnetic field. **c, d,** Corresponding time-reversal process of **a** and **b**. The dotted lines indicate the voltage dependence of the spin-flip magnetic field of the magnetic transitions (Supplementary Section 2 gives more analysis). **e, f,** Critical magnetic fields for the spin-flip transition for forward sweeps with increasing magnetic field as a function of the applied electric field.

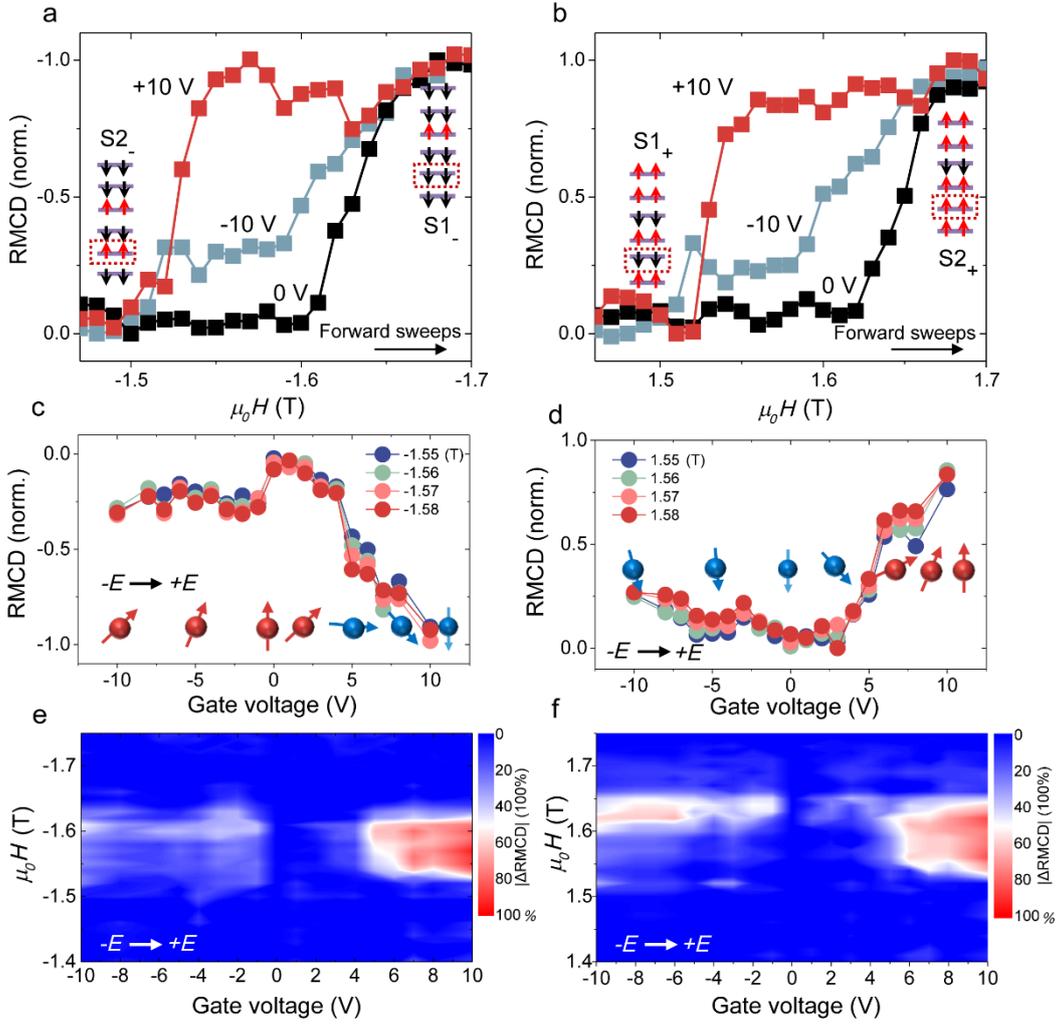

**Fig. 3 | RMCD signal versus applied magnetic field and gate voltage**. **a, b,** Selected horizontal line cuts from Fig. 2b and d demonstrating the gate-induced magnetic transition. **c, d,** Voltage-induced continuous transition from $S1_+$ ($S1_-$) to $S2_+$ ($S2_-$) states at selected magnetic field. A ball with an arrow represents the spin of one layer marked by a red dashed boxes in **a** and **b**, where the arrow represents the spin direction. **e, f,** $\mu_0H$-V dependent difference of the absolute RMCD value, where RMCD as a function of magnetic field at 0 V serves as a base line.

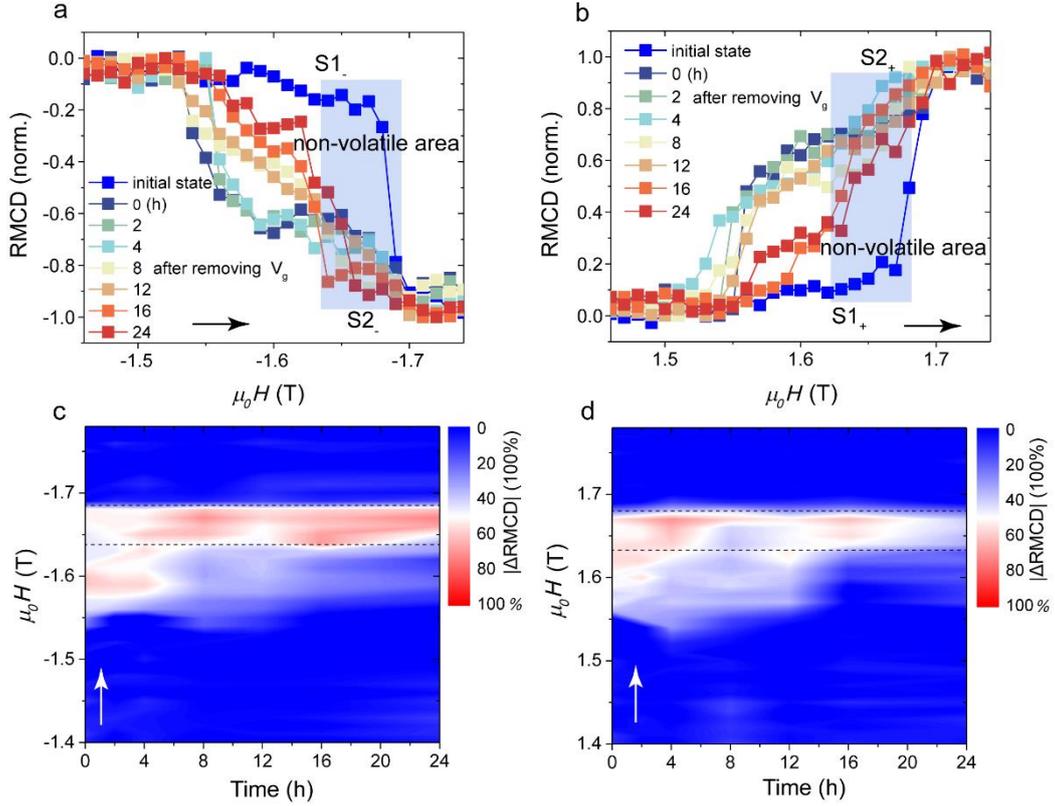

**Fig. 4 | Non-volatile electric control of magnetic transition. a, b,** Time-dependent RMCD intensities as a function of magnetic field after removing applied voltage. The shaded area highlights the non-volatile electric control of magnetic transition. **c, d,** $\mu_0H$-time dependent difference of the absolute RMCD value, where RMCD curve of initial state serves as a base line. Evidently, above 50% of the RMCD intensities has been preserved ranging from +1.64 T (-1.640 T) to +1.68T (-1.68 T) after 24 hours, indicating a non-volatile electric control of magnetic transition (highlight non-volatile area in black dashed boxes). The black dashed boxes in Fig. 4c and d correspond to the shaded areas in Fig. 4a and b. The black and white arrows represent the sweep direction of magnetic field.

14